\documentclass[prl,twocolumn,superscriptaddress]{revtex4}

\usepackage[utf8]{inputenc} 
\usepackage{graphicx} 
\usepackage{hyperref} 
\usepackage{xspace} 

\begin{document}

\title{Magnetic Coupling and Single-Ion Anisotropy\\  in Surface-Supported Mn-based Metal-Organic Networks}

\author{L. Giovanelli}
\affiliation{Aix-Marseille Universit\'{e}, IM2NP, Avenue Escadrille Normandie-Niemen, Case 151, F-13397 Marseille CEDEX 20, France}
\affiliation{CNRS, IM2NP (UMR 6242), Marseille-Toulon, France}

\author{A. Savoyant}
\affiliation{Aix-Marseille Universit\'{e}, IM2NP, Avenue Escadrille Normandie-Niemen, Case 151, F-13397 Marseille CEDEX 20, France}
\affiliation{CNRS, IM2NP (UMR 6242), Marseille-Toulon, France}

\author{M. Abel}
\affiliation{Aix-Marseille Universit\'{e}, IM2NP, Avenue Escadrille Normandie-Niemen, Case 151, F-13397 Marseille CEDEX 20, France}
\affiliation{CNRS, IM2NP (UMR 6242), Marseille-Toulon, France}

\author{F. Maccherozzi}
\affiliation{Diamond Light Source, Didcot, OX11 0DE, United Kingdom}

\author{Y. Ksari}
\affiliation{Aix-Marseille Universit\'{e}, IM2NP, Avenue Escadrille Normandie-Niemen, Case 151, F-13397 Marseille CEDEX 20, France}
\affiliation{CNRS, IM2NP (UMR 6242), Marseille-Toulon, France}

\author{M. Koudia}
\affiliation{Aix-Marseille Universit\'{e}, IM2NP, Avenue Escadrille Normandie-Niemen, Case 151, F-13397 Marseille CEDEX 20, France}
\affiliation{CNRS, IM2NP (UMR 6242), Marseille-Toulon, France}

\author{R. Hayn}
\affiliation{Aix-Marseille Universit\'{e}, IM2NP, Avenue Escadrille Normandie-Niemen, Case 151, F-13397 Marseille CEDEX 20, France}
\affiliation{CNRS, IM2NP (UMR 6242), Marseille-Toulon, France}

\author{F. Choueikani}
\affiliation{Synchrotron SOLEIL, L'orme des Merisiers, Saint-Aubin - BP48, 91192 Gif-sur-Yvette CEDEX, France}

\author{E. Otero}
\affiliation{Synchrotron SOLEIL, L'orme des Merisiers, Saint-Aubin - BP48, 91192 Gif-sur-Yvette CEDEX, France}

\author{P. Ohresser}
\affiliation{Synchrotron SOLEIL, L'orme des Merisiers, Saint-Aubin - BP48, 91192 Gif-sur-Yvette CEDEX, France}

\author{J.-M. Themlin}
\affiliation{Aix-Marseille Universit\'{e}, IM2NP, Avenue Escadrille Normandie-Niemen, Case 151, F-13397 Marseille CEDEX 20, France}
\affiliation{CNRS, IM2NP (UMR 6242), Marseille-Toulon, France}

\author{S. S. Dhesi}
\affiliation{Diamond Light Source, Didcot, OX11 0DE, United Kingdom}

\author{S. Clair}
\affiliation{Aix-Marseille Universit\'{e}, IM2NP, Avenue Escadrille Normandie-Niemen, Case 151, F-13397 Marseille CEDEX 20, France}
\affiliation{CNRS, IM2NP (UMR 6242), Marseille-Toulon, France}

s
\date{\today}

\begin{abstract}

The electronic and magnetic properties of Mn coordinated to 1,2,4,5-tetracyanobenzene (TCNB) in the Mn-TCNB 2D metal-ligand network have been investigated by combining scanning tunneling microscopy and X-ray magnetic circular dichroism (XMCD) performed at low temperature (3 K). When formed on Au(111) and Ag(111) substrate the Mn-TCNB networks display similar geometric structures. Magnetization curves reveal ferromagnetic (FM) coupling of the Mn sites with similar single-ion anisotropy energies, but different coupling constants. Low-temperature XMCD spectra show that the local environment of the Mn centers differs appreciably for the two substrates. Multiplet structure calculations were used to derive the corresponding ligand field parameters confirming an in-plane uniaxial anisotropy. The observed interatomic coupling is discussed in terms of superexchange as well as substrate-mediated magnetic interactions.

\end{abstract}

\maketitle

\section{Introduction}

Exploiting the functionality of organic molecules to manipulate electron spin has been the subject of intense scientific activity during over the last few years \cite{Miyamachi2012,Sanvito2011,Gambardella2009}. The discovery of single molecule magnets displaying high blocking temperatures and quantum tunneling of magnetization suggested an alternative way to  downsizing information storage \cite{Ghigna2001, Schlegel2008, Mannini2009, Stepanow2010b, Miyamachi2012}. At the same time, low-Z organic materials showed high spin-transport coherence properties, making possible the integration of the spin degree of freedom in organics-based semiconductor devices \cite{Raman2013, Sanvito2011}. For both aspects the interface between the magnetically-active constituents has been shown to have a crucial impact \cite{Stepanow2011,Sanvito2010,Javaid2010,Wende2007,Gargiani2013}.

Organic magnets were first reported for large-spin molecules such as $Mn_{12}-ac$ \cite{Thomas1996}. More recently the use of smaller, $\pi$-conjugated macrocycles such as phthalocyanines and porphyrins hosting a single transition metal atom has shown great versatility. This includes, for instance, the possibility to modify the magnetic spin state of the central metal atom through ferromagnetic (FM) coupling to the substrate \cite{Bernien2009} or by further adsorption of smaller molecules \cite{Wackerlin2010, Isvoranu2010}.
$\pi$-conjugated molecules are robust and can incorporate any transition metal. On the other hand they tend to organize via weak intermolecular bonding (van der Waals or H-bonding) thus limiting the possibility of nanostructuration.

Currently an alternative approach for the synthesis of magneto-organic nanostructures is being explored. It consists in manipulating the magnetic properties of transition metal atoms through selective bonding to functional ligands in surface-supported, self-assembled metal-organic networks \cite{Gambardella2009,Umbach2012,Abdurakhmanova2013}. Recent experiments have shown that the magnetic coupling between metal centers as well as their magnetic anisotropy can be controlled by changing the nature of the metal-ligand (M-L) bonding or by adsorption of molecular oxygen \cite{Gambardella2009,Umbach2012,Abdurakhmanova2013}.

Given the precise control offered by the self-assembly approach over the size and shape of the M-L networks, these studies open up a large field of investigation for the development of organic nanostructures with designed magnetic properties. 

Among the key issues to be addressed are the coupling between the metal centers, their magnetic anisotropy, the role played by the M-L interaction and the adsorption interaction. The choice of the metal centers is also important in view of total spin and anisotropy properties. A large anisotropy is necessary for magnetic memory applications whereas a small one opens the way to qubit manipulation \cite{Bertaina2009}.

Mn with five $d$-electrons is a preferred candidate as a spin center in M-L networks. Under the action of organic ligands the presence of intra-molecular exchange interaction and sizable single-ion anisotropy have been reported \cite{Kuepper2011, Prinz2010, Khanra2008} making Mn-based organic molecules suitable for single-molecule magnetism. Recently, detailed structural and electronic studies on Mn-based M-L networks have been reported showing a rich interplay between metal centers, ligands and different substrates \cite{Tseng2009, Faraggi2012}, thus suggesting the possibility of finely tuning the Mn magnetic properties.

The present paper focusses on the electronic structure and magnetic properties of Mn forming a regular M-L network with 1,2,4,5-tetracyanobenzene: Mn-TCNB. This molecule has raised interest recently for its potential ability to undergo a chemical reaction and form phthalocyanine derivatives or polymeric phthalocyanine \cite{Abel2011}. Scanning tunneling microscopy (STM) allowed to the structure of the M-L network grown on two different substrates to be resolved, namely Au(111) and Ag(111). Chemically and magnetically sensitive X-ray magnetic circular dichroism (XMCD \cite{Sthor1999}) performed at low temperature and under variable magnetic field was used to study the effect of the M-L bonding on the Mn magnetic properties. 

The results show that the organic linkers sensibly deform the otherwise spherically symmetric Mn electron cloud. Magnetization curves reveal that the Mn atoms in the M-L network are FM coupled.  Angle-dependent measurements show in-plane uniaxial anisotropy. Ligand-field multiplet calculations allow to reproduce the XMCD spectra obtained on the two substrates. An increased in-plane distortion is found in the case of Ag(111), suggesting a stronger M-L interaction. An analytical expression for the anisotropy energy is derived from the ligand field parameters  confirming the in-plane uniaxial anisotropy. Finally superexchange and substrate-mediated interactions are discussed as possible causes for the observed magnetic coupling.\\

\section{Experiment}

The experiments were performed at the DEIMOS beamline, SOLEIL, France: it is an undulator beamline working in the soft-ray range with variable polarization. The end station is UHV-connected to a sample preparation facility comprising sputtering and annealing for substrate preparation, molecular sublimation, STM and Auger electron spectroscopy (AES). 
Au(111) and Ag(111) surfaces were prepared by repeated cycles of sputtering and annealing. Subsequently a single layer of TCNB was deposited by sublimating the molecules from a crucible while keeping the substrate at room temperature. Finally, additional sublimation of Mn atoms resulted in the formation of the Mn-TCNB M-L network domains with a given stoichiometry. The Au(111) sample was post-annealed at 100$^\circ$ C for 10 minutes to increase the homogeneity of the network. This annealing temperature was chosen to be below the activation temperature of the reaction between TCNB and Mn. Every step of the procedure was monitored by STM and AES. The STM images of Fig.\ref{STM} were obtained at IM2NP in Marseille in equivalent experimental conditions.

X-ray absorption spectroscopy (XAS) with variable polarization was performed in total electron yield. XMCD is the difference between XAS spectra acquired with circularly polarized light, with opposite alignment of X-ray helicity vector (99\% circularly polarized light) and sample magnetization. The spectra were taken at 3 K and under a magnetic field of 6 T applied along the light propagation direction. Measurements were performed at normal incidence (NI, $\Theta=0^\circ$, $\Theta$ being the angle between the surface normal and the light beam) and at grazing incidence (GI, $\Theta=70^\circ$). XMCD was also used to record magnetization curves by scanning the magnetic field and measuring the difference between resonance and off-resonance XAS at each step. The signal was then normalized to the XAS signal at the highest applied magnetic field. The final curves are the average of four magnetic field scans performed after changing alternatively the X-ray helicity and the magnetic field. All the magnetization curves are then normalized for comparison with a model Brillouin function. A least-squared fit to mean field theory spin Hamiltonian containing zero-field splitting and FM coupling terms was performed.\\

\section{Multiplet calculations}

The XMCD spectra calculated in the framework of the ligand-field multiplet with no symmetry constraints. A numerical code was developed which diagonalized the microscopic Hamiltonian in the initial ($2p^63d^5$) and final ($2p^53d^6$) configurations \footnote{The Hartree-Fock energies reduced to 80\% of the atomic values were used as obtained from the CTM4XAS program \cite{Stavitski2010} were used} and then computed the dipole transitions. 
The microscopic Hamiltonian of the $p$ and $d$ many-electron system \
\begin{equation}\label{H}
    H = H_{C} + H_{SO} + H_{LF} + H_{Z}
\end{equation}
contains, respectively, Coulomb repulsion in the $d$-shell and between $p$- and $d$-shell, spin-orbit interaction for $p$- and $d$-electrons, ligand field and Zeeman interaction with an external magnetic field.

Considering a D$_{4h}$ symmetry for the Mn$^{2+}$ environment, the ligand field is defined by $D_q$, $D_s$ and $D_t$ one-electron parameters. The last two give the deviation from a octahedral symmetry, characterized by $D_q>0$. The ligand field Hamiltonian defines the four-fold axis $z$.
The external magnetic field is defined by its magnitude $B$ and the $\theta$ angle it makes with the $z$-axis (within the ($x,z$) plane).

All these parameters being set, $H$ is diagonalized in initial (252 states) and final (1260 states) configurations under saturating magnetic field resulting in a set of eigenvalues and eigenvectors.
Subsequently, the absorption spectra are calculated considering dipole-allowed transitions with circularly-polarized light. The resulting spectra are broadened by a lorentzian function to take into account the finite lifetime of the core-hole.
\\

\section{Results and discussion}

\begin{figure}
\centering
\includegraphics[width=0.48\textwidth] {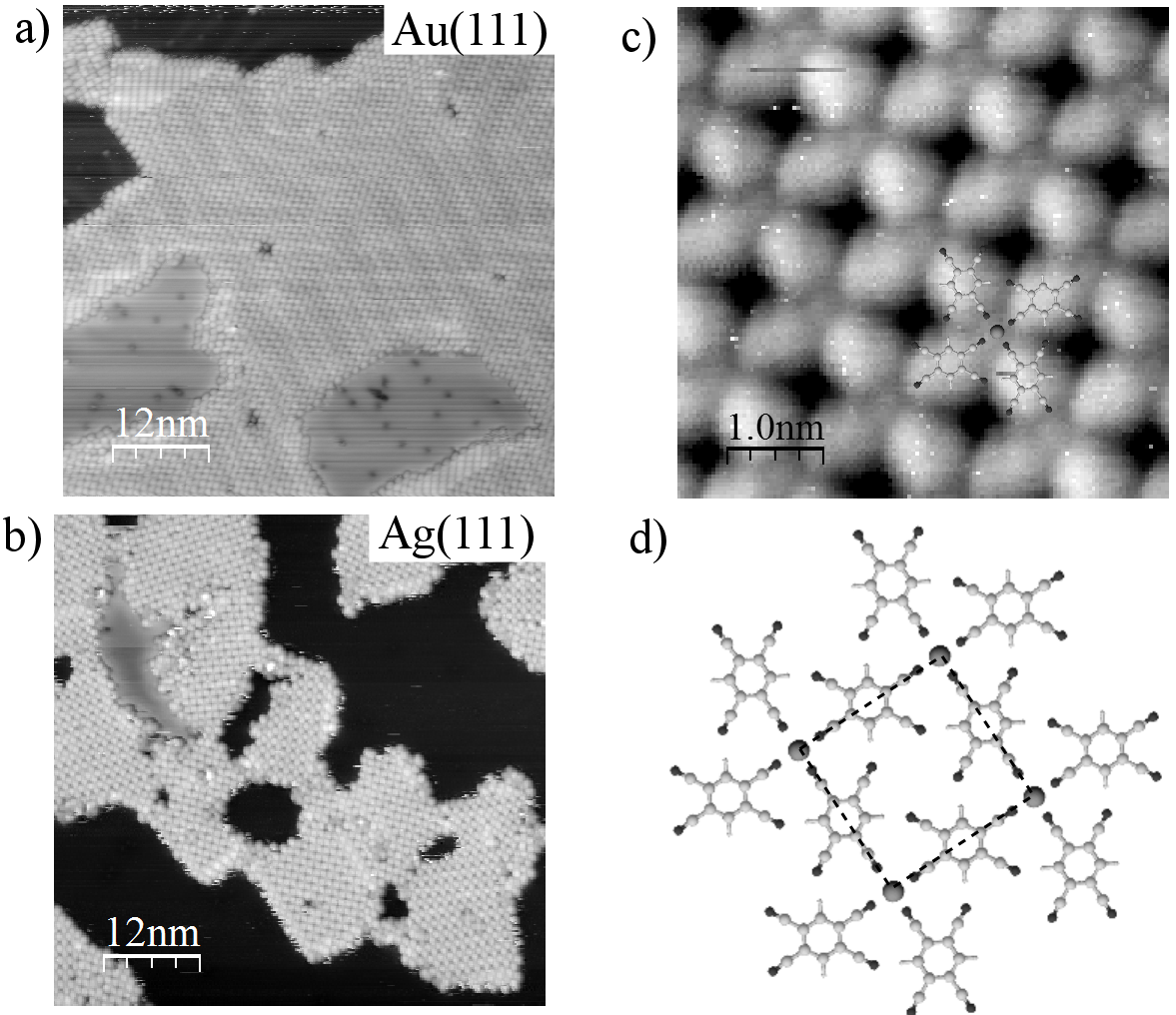} 
\caption{STM images of Mn-TCNB networks formed after deposition on Au(111) (a) and Ag(111) (b). (c) high magnication image of the metal-organic structure on Au(111) with a superimposed schematic model.}
\label{STM}		
\end{figure}

The STM images of Fig. \ref{STM} show the Mn-TCNB networks formed after sequential deposition of TCNB and Mn. On Au(111) the domains are more extended (40 to 50 nm large) than for Ag(111) (about 20 nm large).The unit cell is square with a 1.2$\pm$0.1 nm periodicity and comprises two molecules and one metal atom for both substrates thus representing a Mn(TCNB)$_2$ stoichiometry. Fig. \ref{STM}-c presents a high magnification image of the metal-organic structure with a superimposed schematic model, showing that each Mn atom is linked to 4 TCNB molecules through a Mn-N metal-ligand bonding displaying $D_{4h}$ symmetry.

\begin{figure}
\centering
\includegraphics[width=0.4\textwidth] {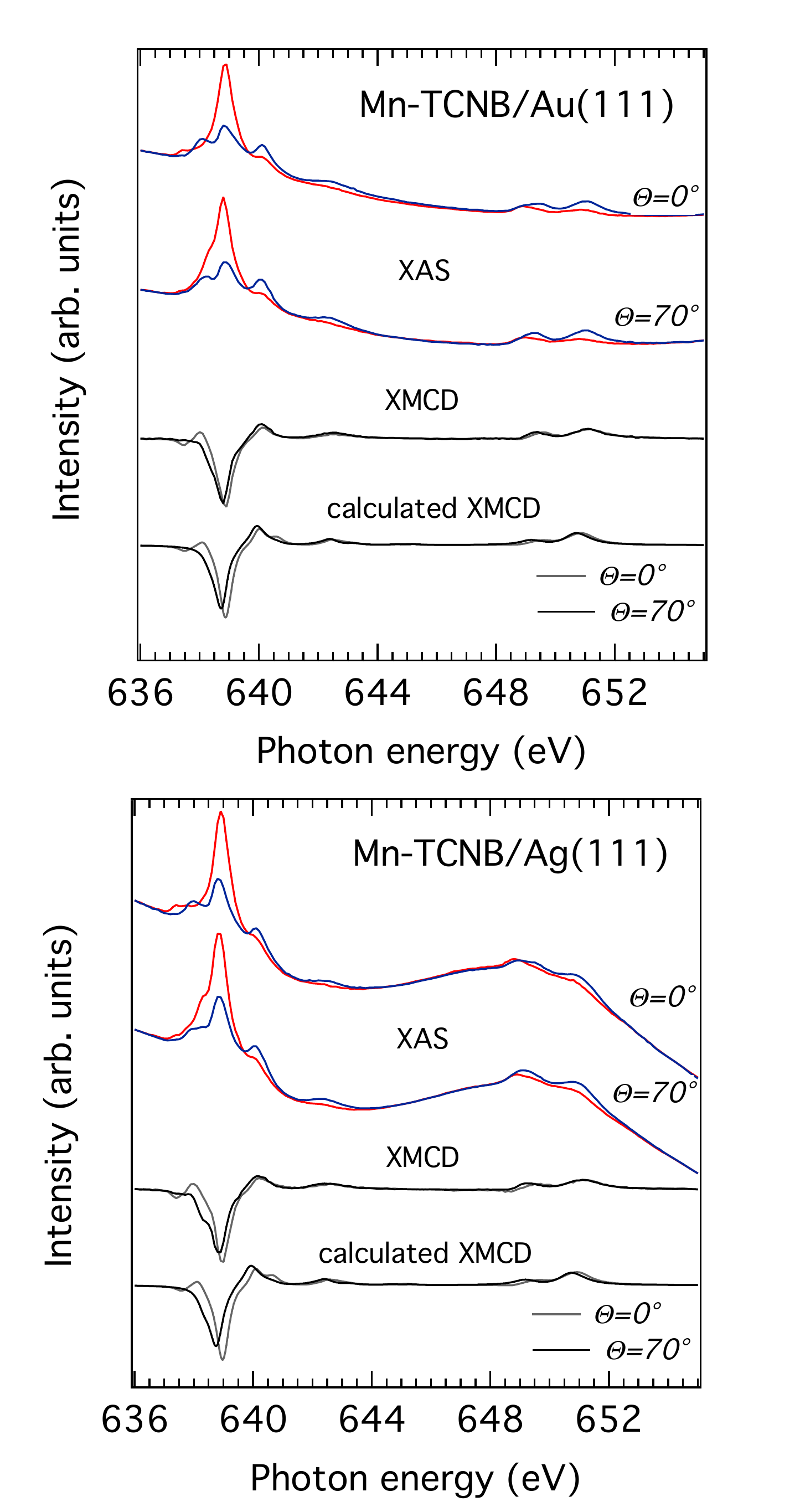} 
\caption{Angular dependence of the XAS and resulting XMCD over the Mn $L_{2,3}$ edge for (a) Mn-TCNB/Au(111) and (b) Mn-TCNB/Ag(111) measured at $3 K$  with an applied magnetic field of $6 T$ parallel (blue) and antiparallel (red) to the X-ray beam helicity. Two sample orientations with respect to the light propagation direction were measured ($\Theta=0^\circ$ correspond to normal incidence (NI) and $\Theta=70^\circ$ to grazing incidence (GI)). The bottom curves are obtained by ligand-field multiplet calculations (see text). A normalization factor of 1.4 and 2.0 was applied to the calculated XMCD for Au(111) and Ag(111) respectively.}
\label{xas_xmcd}		
\end{figure}

In Fig. \ref{xas_xmcd} the XAS and XMCD spectra over the Mn $L_{3,2}$ edge are displayed for the two samples with the X-ray beam at different incidence angles. Despite the differences due to substrate background contributions the overall shape of the XAS spectra are similar for the two systems: both exhibit the spectral features typical of Mn$^{2+}$ with a $d^5$ configuration.\cite{Kuepper2011,Prinz2010,Khanra2008,Arrio1996, Nanba2012, Gambardella2009, Gambardella2007, Nagel2007, Kang2008, Edmonds2006, Maccherozzi2008, Cramer1991, DeGroot1990}. For both systems a clear difference is observed at the $L_3$ in going from NI to GI indicating a preferred orbital orientation. 

The XMCD also displays anisotropy for both systems, more markedly at the $L_3$. In NI a negative-positive pre-edge is followed by a strong but featureless negative peak (at 638.9 and 639 eV for Au and Ag respectively). In GI there is no clear pre-edge feature, but the main negative peak (638.8 and 638.9 eV for Au and Ag respectively) has a shoulder at low energy side. Finally the main peak intensity is weaker at GI than at NI. At the $L_2$ the differences are smaller but it can be seen that the onset of the white line is shifted to lower energies in GI.

The XMCD spectra changes significantly in going from Au(111) to Ag(111) substrate. The first difference is the quoted shift of 0.1 eV to higher energy for both sample orientations. The second, more clearly noticeable aspect, is that on Ag(111) the shoulder at the main negative peak in GI is more pronounced than on Au(111) and the GI to NI difference of peak height is more marked. Finally the onset shift at the $L_2$ is larger for Ag(111).

\begin{figure}
\centering
\includegraphics[width=0.4\textwidth] {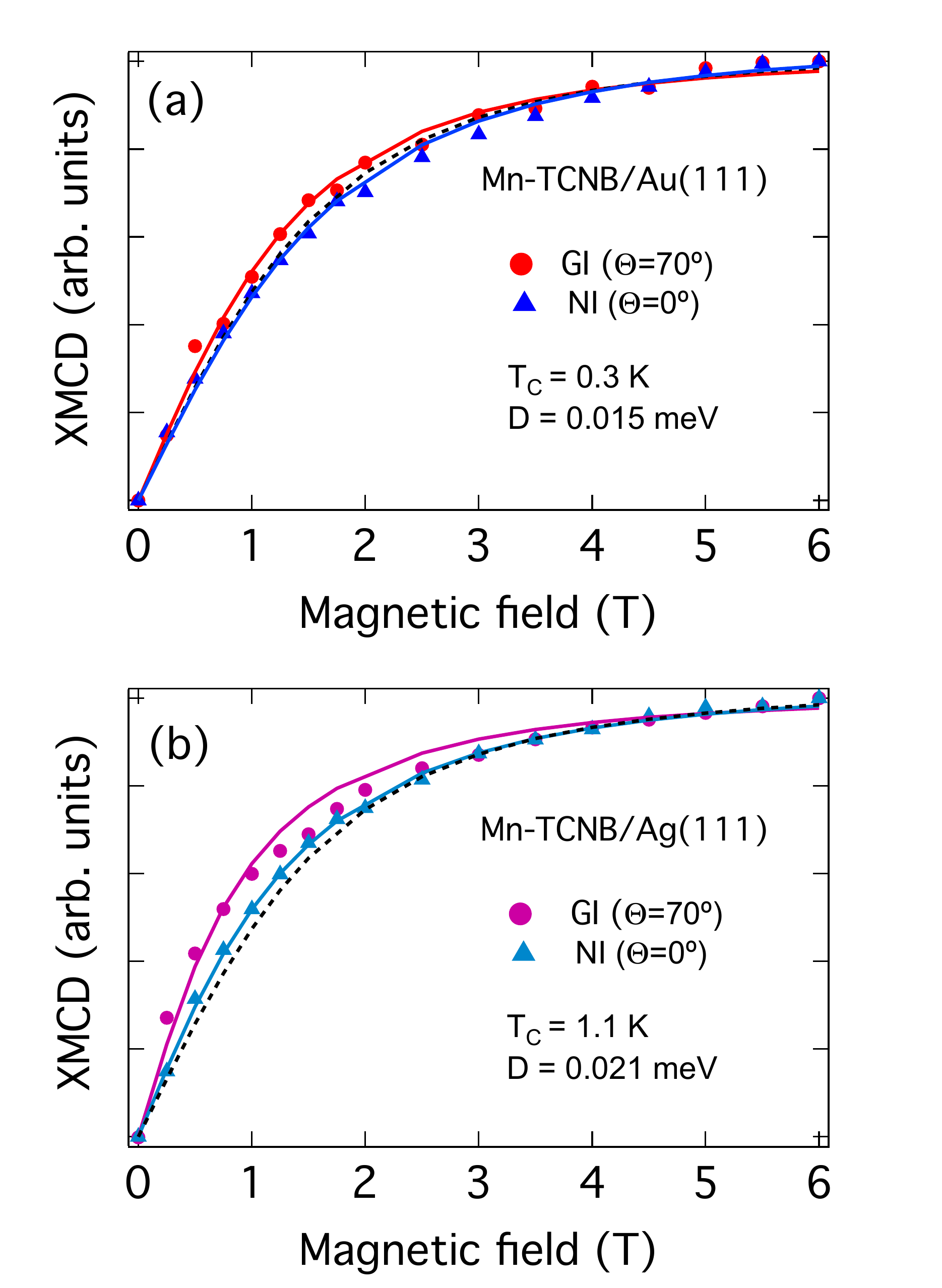} 
\caption{Magnetization curves (symbols) as obtained from XMCD of (a) Mn-TCNB/Au(111) and (b) Mn-TCNB/Ag(111) at $3 K$ in NI and GI geometries. Superimposed (continuous lines) are the least-squared fitting curves obtained through equation (\ref{H_spin}) (see text for details). The resulting fitting parameters for the two systems are indicated. The Brillouin function at $3 K$ is shown for comparison as a dashed line.}
\label{M_vs_B}		
\end{figure}

The magnetization curves for the two systems probed along the surface normal (NI) and close to sample surface plane (GI) are displayed in Fig. \ref{M_vs_B}. For Mn-TCNB/Au(111) a small in-plane magnetic anisotropy is detected. Both curves are close to the Brillouin function for a $S=5/2$ system at $3$ K. The same kind of anisotropy is measured for Mn-TCNB adsorbed on Ag(111). Remarkably, for Ag(111) a clear departure from the Brillouin function is observed suggesting a FM coupling \cite{Umbach2012, Abdurakhmanova2013}. The anisotropy observed for both systems indicates an easy plane parallel to the network plane.

In order to get more insight in the physical parameters governing the shape of the measured curves, we used a spin Hamiltonian with zero-field splitting (parameter $D$) and magnetic coupling between individual spins. This was done in the framework of the mean field theory, where the magnetic interaction of a spin with all other spins is replaced by an effective mean magnetic field $\bf{B_{eff}}$ added to the external applied field $\bf{B_{ext}}$ and proportional to the mean magnetization $\bf{S}$
\begin{equation}\label{H_spin}
	H_S=DS^{2}_{z}-g\mu_{B}\bf{B\cdot S}
\end{equation}

where $\bf{B=B_{ext} + B_{eff}}$ and $\bf{B_{eff}}=\lambda \bf{S}$.
The Curie temperature associated with the magnetic coupling is given by $T_{c}=\lambda S(S+1) g \mu_{B} / (3 k_{B})$ \cite{Ashcroft_Mermin}. The strength of the coupling is related to the Curie temperature  as $T_c=\sum\nolimits_{i}J_i/(3k_B)$, where the sum goes over the 4 first neighboring sites of the square Mn-lattice.
In fact the magnetic field includes also a contribution of the local field created by dipolar interaction of all surrounding spins:
\begin{equation}\label{B}
	\bf{B = B_{ext} + B_{eff} + B_{dipol}}
\end{equation}

For an infinite 2D square spin lattice of size $a_{0}$, $\bf{B_{dipol}}$ is of the order of $(8 \pi/3a_{0}^3) g \mu_{B}  \bf{S}$ \cite{Yafet1988} and is about one order of magnitude smaller than the effective field $\bf{B_{eff}}$ due to magnetic interaction. Note that both terms, zero-field splitting and magnetic interaction are necessary to obtain a reasonable fit to the experimental data.

The experimental curves are fitted by the model with the following parameters (see supplementary information for details): $D$=0.040 meV and $T_c$=0.82 K ($J_i=0.05 meV$) for Au(111); $D$=0.035 meV and $T_c$=1.62 K ($J_i= 0.10meV$) for Ag(111). The sign of the anisotropy parameter indicates an easy-plane uniaxial anisotropy for both systems. The anisotropy energy is similar but the coupling is about twice as strong for the Ag(111) substrate.

The zero-field splitting (or single-ion anisotropy) energy results from the combined effect of the ligand field acting on the Mn atoms and the atomic $d-d$ spin-orbit interaction. Because these parameters affect the XMCD their approximate values can be inferred by comparing the experiment to parameter-dependent model spectra.
XMCD spectra with varying ligand field parameters were therefore calculated. The best agreement was obtained when 10$D_q$=0.7 and $D_t$=0.07 $eV$ for both substrates and $D_s$=0.07 and $D_s$=0.10 $eV$ for Au(111) and Ag(111), respectively. The calculated spectra are displayed in the bottom parts of Figs. \ref{xas_xmcd}(a) and \ref{xas_xmcd}(b). The overall agreement of the simulations is satisfactory. All main experimental features are well reproduced in terms of energy and intensity. Namely the negative-positive pre-edge is present in the NI but absent in GI where a single negative peak is found instead. When Au(111) is changed to Ag(111) a higher $D_s$ is found. This has several effects on the simulated spectra: $(i)$ in GI a shoulder to the main negative peak develops at low energy; $(ii)$ the difference in intensity at the main peak between GI and NI increases; $(iii)$ at the $L_2$ the onset shift becomes larger. All these features allow to correctly reproduce the main differences between the experimental spectra relative to Au and Ag substrates.

Despite the very good line shape agreement, the experimental XMCD is sensibly smaller than expected (a normalization factor of 1.4 and 2.0 was applied to the calculated XMCD for Au(111) and Ag(111), respectively). Similar reductions were found in other studies of Mn impurities on surfaces \cite{Gambardella2005, Durr1997}. On Ge and GaAs that was ascribed to a reduction of the spin moment due to hybridization of Mn $d$-electrons with substrate states. Such hybridization can be ruled out in the present case since the XMCD spectral shape does not show the expected broadening \cite{Gambardella2005}. When deposited on FM substrates sub-ML quantities of Mn show a reduced XMCD due to the Mn-Mn bond formation resulting in $d-d$ overlapping and consequent AFM coupling \cite{Durr1997}. In the present case it is indeed possible that part of the Mn atoms are engaged in AFM coupling thus reducing the relative intensity of the XMCD signal. Another possible explanation may reside in the interaction of Mn $d$-states with delocalized substrate $s$-electrons. Such coupling was recently proposed to explain a quenching of the magnetic moment in CoPc/Au(111) \cite{Stepanow2011}. This scenario is indeed appealing but needs to be studied by interface charge transfer multiplet calculation, an approach that is beyond the scope of the present paper.
 
Even without such refinements the ligand field parameter sets can be used to calculate $D$ by means of the microscopic Hamiltonian. This was done analytically by exactly solving  the Coulomb operator $H_C$ in the initial configuration ($\equiv$ $d^5$) and then treating $H_{SO}$ and $H_{LF}$ as perturbations to the ground sextet ($^6S$) energy. The results is to split this sextet into three doublets ($|m_S \rangle$ = $|\pm 5/2 \rangle$, $|\pm 3/2 \rangle$, $|\pm 1/2 \rangle$ from highest to lowest lying) separated by $4D$ and $2D$, respectively. Similarly to what is obtained for Mn$^{2+}$ in C$_{3v}$ symmetry \cite{Savoyant2009}, the fourth order expression of axial anisotropy for a $d^5$ configuration in D$_{4h}$ symmetry is derived  \footnote{When calculating the energies of $H=DS_{z}^{2}$ for a spin 5/2 (six states) one finds 3 doublets separated by 2D and 4D (giving 6D between the lowest and the highest). The perturbative calculation of level $^{6}S$  gives exactly 3 doublets separated by the same ratio, which allows to assign them to those of an effective 5/2 spin under the action of $H=D S_z^2$. It should be noticed that the expression for $D$ is equally valid for the following  symmetries: D$_4$, D$_{2d}$ and C$_{4v}$} :

\begin{equation}\label{D}
    D^{(4)} = \frac{63}{10} \frac{\zeta_d^2D_s}{\mathcal{P}^2\mathcal{D}}(\zeta_d-\frac{D_s}{3}) + 35 \frac{\zeta_d^2D_t}{\mathcal{P}^2\mathcal{G}}(2D_q-D_t)
\end{equation}

where $\mathcal{P}=7\mathcal{(B+C)}$, $\mathcal{D}=17\mathcal{B}+5\mathcal{C}$ and $\mathcal{G}=10\mathcal{B}+5\mathcal{C}$ are, respectively, the distance between ground sextet $^6S$ and excited quartets of the $d^5$ configuration lying just above in energy: $^4P$, $^4D$ and $^4G$. $\mathcal{B}$ and $\mathcal{C}$ are the Racah parameters, related to the Slater-Condon's by $F_{dd}^2 = 49\mathcal{B} + 7\mathcal{C}$ and $F_{dd}^4 = 63\mathcal{C}/5$. 
This formula relates one-electron ligand-field (electric field plus hybridization) anisotropy parameters ($D_q, D_t$ and $D_s$) to the magnetic anisotropy parameter ($D$) of a $S=5/2$. (\ref{D}) is thus very important to predict the effect of a modified chemical environment on the magnetic anisotropy properties.

When using $\zeta_d=0.052$ eV, the resulting value of $D$ is 0.012 meV and 0.011 meV for Au and Ag respectively. Such values are somehow smaller than those obtained by fitting the magnetization curves (0.040 meV and 0.035 meV, respectively). Nevertheless they confirm the presence of the single ion, easy-plane, uniaxial anisotropy \footnote{It has to be stressed out that values of $D$ closer to what obtained from fitting the magnetization curves can be obtained through (\ref{D}) by further increasing the spin-orbit parameter $\zeta$ and without altering the simulated XMCD spectrum appreciably. Nevertheless, a free ion value was chosen since $\zeta$ is known to be \emph{reduced} by the effect of the environment}.
Moreover, from (\ref{D}) it results that although the in-plane tetragonal distortion $D_s$ affects the angular dependence of the XMCD spectrum (of course in combination with $D_t$), it only has a weak influence on the anisotropy energy which is determined mainly by $D_q$ and $D_t$ through the spin-orbit interaction and is therefore very similar for the two substrates.\\


The results reported above indicate that within the Mn-TCNB M-L network, magnetic anisotropy is induced by the joint effect of ligand field and spin-orbit interaction. Comparatively, the zero-field spitting reported recently for high-spin, $d^5$ Mn in a star-shaped heteronuclear complex ($Cr^{III}Mn^{II}_3$) is as high as 0.124 meV \cite{Prinz2010}.
On the other hand the magnetic coupling constant found here is higher than reported for star-shaped molecules having shorter Mn-Mn distance \cite{Prinz2010, Khanra2008}. 
This may be due to the extended two-dimensional character of the present system and to the presence of the metallic substrate, favoring delocalization of magnetic excitations (see also Refs. \cite{Umbach2012,Abdurakhmanova2013}).

In recent studies on 2D self-assembled M-L networks on noble metals FM behavior was measured \cite{Umbach2012,Abdurakhmanova2013}. In both cases a superexchange mechanism was suggested. For Fe-T4PT/Au(111) the spin-density oscillation across the ligands would favorably propagate through the network \cite{Umbach2012,Bellini2011}. In the case of Ni-TCNQ networks FM is observed for deposition on Au(111), but not on Ag(100). This is explained in terms of different charge transfer channels in the Ni-TCNQ bonding \cite{Abdurakhmanova2013}. In both studies the Ruderman-Kittel-Kasuda-Yosida (RKKY) interaction was considered to be unlikely to explain the FM behavior.

For Mn-TCNB the number of linker atoms between each Mn is even and thus a superexchange interaction through spin density oscillation should be AFM \cite{Bellini2011}. A charge-state dependent coupling as observed for Ni-TCNQ is not observed for Mn-TCNB where XAS spectra reveal that the charge state of the Mn atoms is the same on both substrates. Nevertheless one cannot rule out superexchange as a driving interaction in the case of Mn-TCNB. In such a perspective it should be noticed that for the Ag-supported system the $D_s$ obtained from the XMCD simulation is higher than for Au. This may arise from a more effective M-L linkage resulting in a stronger magnetic coupling.

RKKY interactions could also contribute to the observed FM coupling \cite{Tsukahara2011} in Mn-TCNB. In fact, above (111) noble metal surfaces collective screening occurs through the Schockley state and, to a first approximation, the key parameter determining the sign of the exchange interaction is the product of the lattice constant and the Fermi wave vector ($k_F$). Using the expression for RKKY interactions in 2D \cite{Fischer1975} one observes that the relatively small $k_F$ of Ag(111) favors FM coupling between nearest neighbors (placed 1.2 nm apart). On the other hand, for Au(111) the two Rashba-splitted Schockley states have larger $k_F$ values and at 1.2 nm distance the first minimum for the RKKY interaction is reached, resulting in a moderate AFM coupling. Certainly the real situation is more complex than the above scenario. Nevertheless, in a picture in which RKKY would compete with other channels of magnetic coupling (such as superexchange as mentioned earlier), the FM to AFM screening in going from Ag(111) to Au(111) may help to explain the differences measured in the magnetization curves of the two otherwise very similar Mn-TCNB networks. It would be interesting to test such a scenario for Ag(100) substrate where the surface states are unoccupied and no long-range oscillation of the exchange interactions are expected \cite{Simon2011}.\\

\section{Conclusion}

In summary, the structure and the magnetic properties of two Mn-TCNB metal-ligand networks were studied at low temperature by XMCD. Angle-dependent magnetization curves show FM coupling between the equally spaced Mn atoms with in-plane uniaxial anisotropy. Ligand-field multiplet calculations were used to simulate the experimental spectra. The obtained parameter values allow to estimate the anisotropy energy, confirming the magnetization curves analysis. The spectroscopic differences between Mn-TCNB on Au(111) and Ag(111) indicate a stronger in-plane distortion of the ligand-field for adsorption on Ag(111). This may be the result of a stronger metal-ligand interaction favoring superexchange coupling. Another possible explanation for the different magnetic coupling between Ag(111) and Au(111) is given in terms of RKKY interaction. Because M-L networks are versatile extended 2D systems in which the inter-atomic distance is controlled by the choice of the organic linkers, in the future they may emerge as a new approach to the study of surface magnetic screening phenomena beside single atom manipulation \cite{Khajetoorians2012} and self assembly of $\pi$-conjugated molecules containing magnetic centers \cite{Tsukahara2011}. Finally, an expression (\ref{D}) is given for the magnetic anisotropy energy of a $d^5$ configuration in a $D_{4h}$ environment as a function of crystal field parameters. This should be seen as helpful tool to predict, and possibly tune, the magnetic anisotropy properties via organic linkers. A future development of the model will  focus on the magnetic coupling, possibly including hybridization and surface-mediated interactions.\\

\section{Acknowledgments}
The preparation chambers of the DEIMOS beam line have been partially funded by the Agence National de la Recherche (grant ANR-05-NANO-073).\\

\section{Supplementary information}

\begin{figure}
\centering
\includegraphics[width=0.48\textwidth] {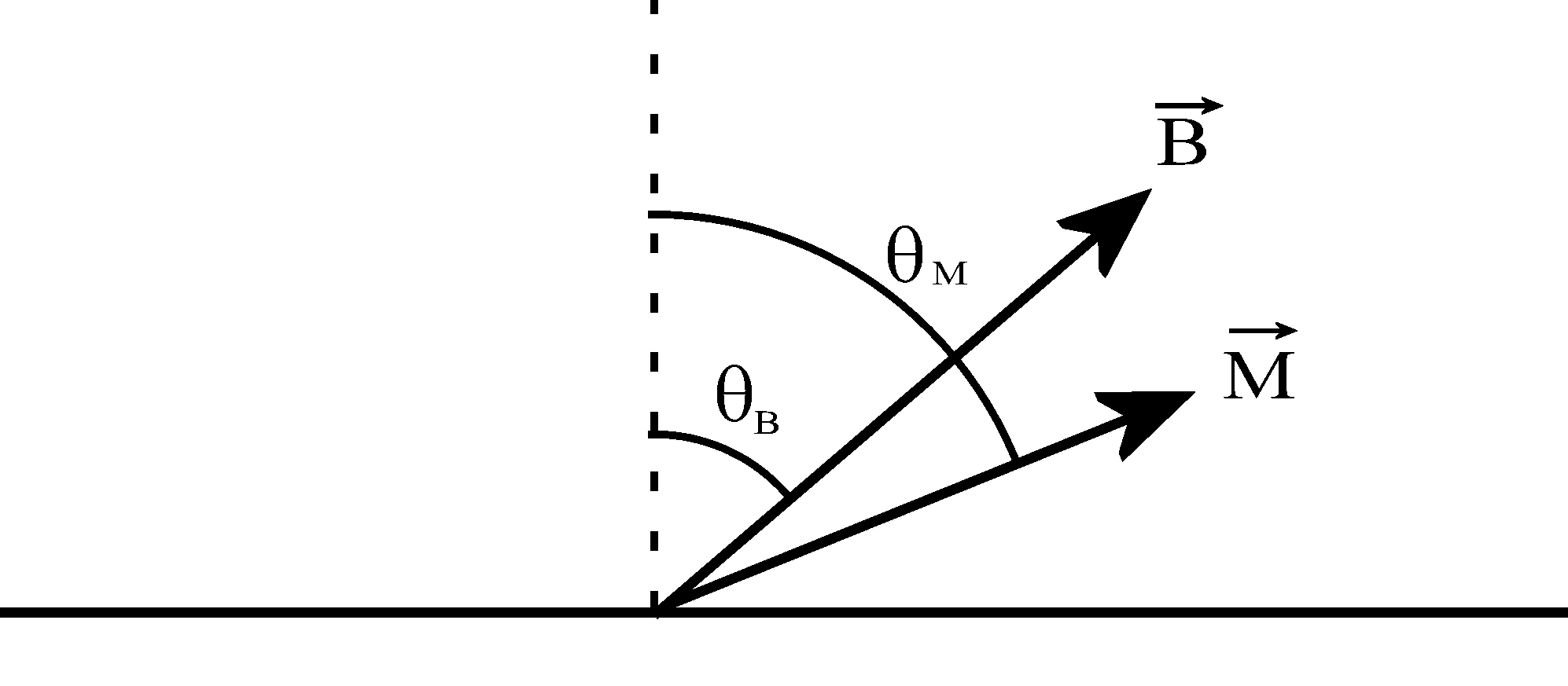} 
\caption{Schematics of relative orientations of $\bf{B}$ and $\bf{M}$. The angles are relative to the surface normal (magnetic hard axis). The direction of the X-ray beam is aligned with the applied magnetic field.}
\label{Schema_XMCD}		
\end{figure}

The spin Hamiltonian (\ref{H_spin}) was solved to calculate the average magnetization vector \textbf{M} depending on the applied field \textbf{B}. However, due to the zero-field splitting term, \textbf{M} and \textbf{B} are in general not aligned (except when \textbf{B} is applied along the easy-plane).
The experimental XMCD signal is only sensitive to the projection of \textbf{M} along the direction of the X-rays (which is in our case identical to the direction of the applied field \textbf{B}, see Fig. S\ref{Schema_XMCD}). Furthermore, it also accounts for a magnetic dipole operator term, the angle dependence of which can be approximated to $(1 - 3 \cos(2 \Theta_M))$ \cite{Stepanow2010,Umbach2012} (Fig. S\ref{Schema_XMCD}). As a result, the XMCD intensity was calculated as
\begin{equation}\label{I_XMCD}
I_{XMCD}=A \cos(\Theta_M - \Theta_B) |M| (1-C(1-3 \cos(2 \Theta_M)))
\end{equation}

\begin{figure}
\centering
\includegraphics[width=0.48\textwidth] {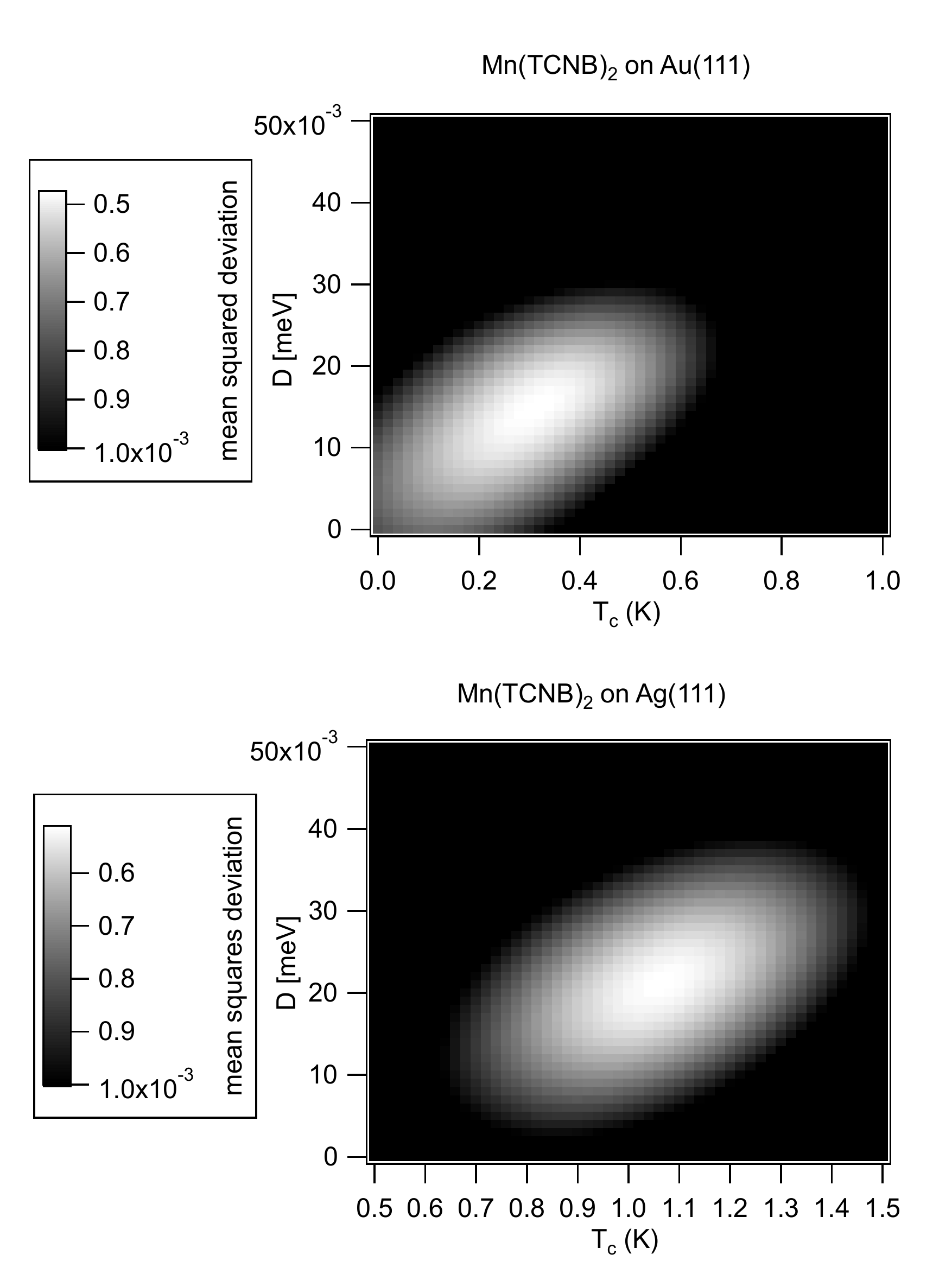} 
\caption{Mean squared deviation of the simulated magnetization curves with respect to the experimental data calculated for the two substrates. The black regions correspond to deviations $\geq 10^3$. The best fit is obtained with D=0.040 meV and Tc=0.62 K for Au(111), and D=0.035 meV and Tc=1.82 K for Ag(111)}
\label{least_square_fit}		
\end{figure}

For every set of $D$ and $T_c$, the parameters $A$ and $C$ are simultaneously tuned to minimize the total mean squared deviation for grazing and normal incidence of the calculated curves with respect to the experimental data. The resulting deviations are reported in Fig. S\ref{least_square_fit} for the Au(111) and Ag(111) substrates. In both cases, a clear minimum is found. 

\bibliography{Ref_Deimos_2012_b}

\end{document}